\documentclass[a4paper,12pt]{article}

\usepackage[pdftex]{graphicx}
\usepackage{latexsym,amsfonts,amsmath,amscd,epsfig}
\usepackage{amssymb,amsxtra,mathtools}
\usepackage{url}
\usepackage{pdflscape}
\usepackage[cp1251]{inputenc}
\newcommand{\bec}{\begin{center}}
\newcommand{\ec}{\end{center}}
\newcommand{\bee}{\begin{equation}}
\newcommand{\ee}{\end{equation}}

\DeclareMathAlphabet{\mathbi}{OML}{cmm}{b}{it}
\title{ The study of the extended Higgs boson sector within 2HDM model}

\author{T.V. Obikhod\thanks{E-mail: obikhod@kinr.kiev.ua}, E.A. Petrenko\\
\small\emph{Institute for Nuclear Research, National Academy of Science of Ukraine} \\
\small\emph{47, prosp. Nauki, Kiev, 03028, Ukraine}}

\date{\small\today}
\begin{document}
\maketitle

\abstract{Consideration of the latest experimental data on the searches for extended sector of Higgs bosons produced at the LHC at a center-of-mass energy of 13 TeV, allows for computer modeling of the properties of supersymmetric particles within 2HDM model. The experimental restrictions on model parameters accounted in FeynHiggs code that is implemented in SusHi program, gave us the possibility to calculate the cross sections and branching fractions for three mechanisms of production and decay of Higgs bosons: 1) pp$\rightarrow H\rightarrow\tau\tau$, 2) pp$\rightarrow A\rightarrow Zh\rightarrow llbb$, 3) pp$\rightarrow H\rightarrow hh\rightarrow bb\tau\tau$  at a center-of-mass energy of 14 TeV. The considered computer modelling make it possible to draw conclusions about the need to take into account the b-associated production process of Higgs bosons for fermionic decay channel at large values of tan$\beta$. Differential cross sections with respect to the Higgs transverse momentum $p_t$ and pseudorapidity $\eta$ are calculated and the peculiarities of the kinematics of the Higgs boson decay products are recognized.}\\
\vspace*{3mm}




\newpage
\section{Introduction}
The Higgs boson, which appears in the models of the spontaneous breaking of electroweak symmetry and is responsible for the occurrence of the masses of elementary particles was discovered on July 4, 2012 at the LHC \cite{1.}. This particle was observed in pp collisions, mainly as a result of the gluon-gluon fusion, and its search is performed in almost all possible decay channels: W and Z bosons (WW and ZZ), bottom quarks (bb), $\tau$ and $\mu$ leptons ($\tau\tau, \mu\mu$), photons ($\gamma\gamma$). 

Search for the Higgs boson is based on the comparison of experimental measurements with theoretical predictions of the Standard Model (SM). The detailed study of the production and decay modes of the new particle with mass of 125–126 GeV at the LHC indicates that the new particle is indeed compatible with the SM Higgs boson. Nevertheless, many scenarios of physics beyond SM include a SM-like Higgs boson as part of an extended sector of scalar particles. In any case, searches for new Higgs bosons are connected with the measurements of the properties of the new particles of an extended sector.

In this aspect, it is necessary to pay attention to the problem of the radiative corrections to the mass of the Higgs boson, the solution of which is associated with the introduction of new particle, so-called superparticle presented in Fig. 1\\
\bec
{\includegraphics[width=0.45\textwidth]{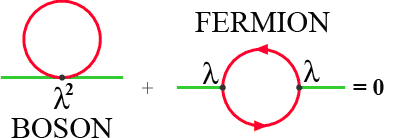}}\\
\emph{{Fig.1.}} {\emph{Presentation of hierarchy problem solution.}}\\
\ec

After mass renormalization between fermionic quark loop and scalar squark loop,  the Higgs boson quadratic mass is limited
\[\Delta m_H^2=\frac{\lambda^2_f}{8\pi^2}\Biggl[6m^2_f\mbox{ln}\frac{\Lambda}{m_f}
-2m^2_S\mbox{ln}\frac{\Lambda}{m_S}\Biggr]\ ,\]
where $m_f$ and $m_S$ are masses of fermion and its superparticle, $\lambda_f$ is Yukawa coupling,
$\Lambda$ is the scale up to which the SM is valid.

The limitation of SM is illustrated through the renormalization-group behavior of Higgs self-coupling $\lambda$. It depends on the numerical values of the SM parameters and defines the Landau pole to the scale of $10^{19}$ GeV. This means that there must be a new physics at energies that are significantly lower the Planck scale \cite{2.}. Such behavior of self-coupling constant which depends also on other parameters (masses of the top quark, $M_t$, of $Z$ boson, $M_Z$, of the Higgs boson, $M_h$, and on the strong coupling constant, $\alpha_s$) 
\[\lambda(M_{Pl})=-0.0143-0.0066\Biggl(\frac{M_t}{\mbox{GeV}}-173.34\Biggr)+\]
\[+0.0018\frac{\alpha_s(M_Z)-0.1184}{0.0007}+0.0029\Biggl(\frac{M_h}{\mbox{GeV}-125.15}\Biggr)\]
creates the problem of electroweak vacuum instability.
The fact of possible existence of new physics at the TeV scale can be studied in deviations of the Higgs self-coupling constant from SM in the process of Higgs boson formation and decay. 
The signal strength $\mu$ for the production $\mu_i$ and decay $\mu^f$ mode of Higgs boson 
$i\rightarrow H\rightarrow f$ is defined as
\[\mu_i=\frac{\sigma_i}{(\sigma_i)_{SM}} \ \mbox{and}\ \ \mu^f=\frac{B^f}{(B^f)_{SM}} \ , \ \] 
where $\sigma_i$ ($i=ggF$ (Gluon fusion), VBF (Vector boson fusion), $WH$ and $ZH$ (Higgs Strahlung), $ttH$ (Top fusion)) and $B^f$ ($f= ZZ,\ WW,\ \gamma\gamma,\ \tau\tau,\ bb,\ \mu\mu$)
are respectively the  production cross sections and the decay branching fractions for Higgs boson. The combined results of the ATLAS and CMS collaborations of Higgs boson production processes $\mu_{ggF}, \mu_{VBF}, \mu_{WH}, \mu_{ZH}, \mu_{ttH}$ and decay signal strengths $\mu^{\gamma\gamma}, \mu^{ZZ}, \mu^{WW}, \mu^{\tau\tau}, \mu^{bb}$ for the combined $\sqrt{7}$ and $\sqrt{8}$ TeV data are presented in Fig. 2.\\
\bec
{\includegraphics[width=0.82\textwidth]{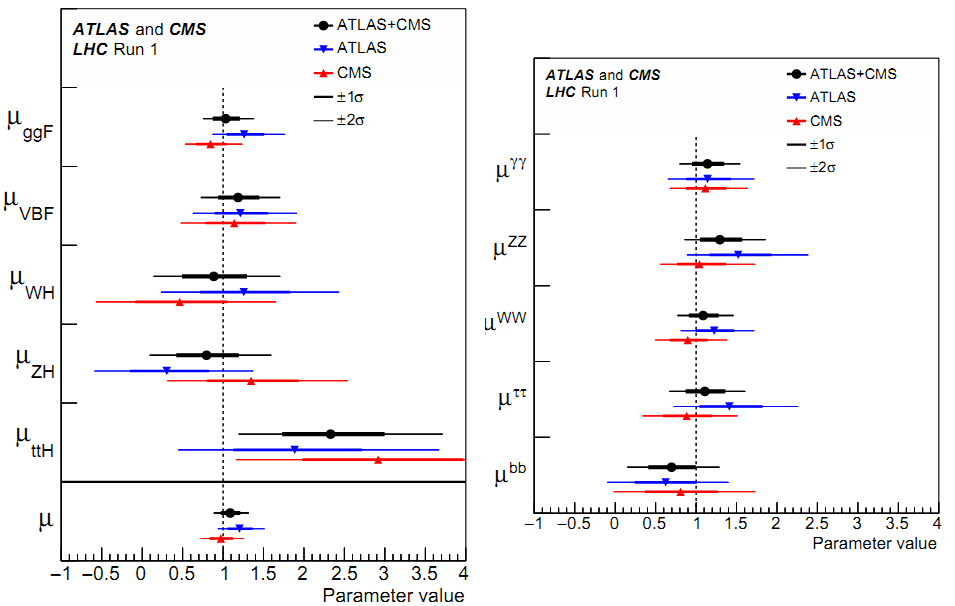}}\\
\emph{{Fig.2.}} {\emph{Production signal strengths (left)and decay signal strengths 
(right) for the combination of ATLAS and CMS Collaborations from \cite{3.}}}\\
\ec

The results of experimental measurements show deviations from the SM.
	
Thus, the existance of Landau pole, the problem of electroweak vacuum instability as well as the experimental data on the production and decay signal strengths of Higgs boson, tiny Higgs mass protected from quantum corrections, prove the necessity of searches for new physics beyond the SM. 

One of the models of beyond the SM physics is the two-Higgs doublet (2HDM) model \cite{4.}. This model provides a solution to the hierarchy problem and predicts five Higgs bosons: two neutral CP-even Higgs bosons, $h, H$ , one neutral CP-odd Higgs, $A$ and two charged Higgs bosons, $H^{\pm}$. Higgs sector of this model can be represented by two free parameters: the mass of the pseudoscalar Higgs boson, $M_A$, and the ratio of the vacuum expectation values of the two Higgs doublets of Higgs sector, tan$\beta$. 

The searches for evidence of beyond SM Higgs bosons is an integral part of Run II at the LHC with the center-of-mass energy of 13 TeV. 
Experimental searches for Higgs sector were performed at the LHC (CMS) \cite{5.} according to the following decay channels, presented in Fig. 3 \vspace*{3mm} \\
\bec
{\includegraphics[width=0.65\textwidth]{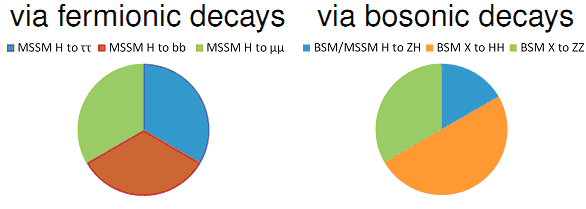}}\\
\emph{{Fig.3.}} {\emph{Decays of the Higgs boson via fermionic channel (left) and bosonic channel (right).}}\\
\ec

In this paper we will consider the following decay channels of beyond SM Higgs bosons:\\
1)	via fermionic decays \\
$\bullet$ $H\rightarrow \tau\tau$\\
2)	via bosonic decays \\
$\bullet$	$A\rightarrow Zh$ \\
$\bullet$	$H\rightarrow hh$ .\\
With the help of SusHi code we will study the properties of beyond SM Higgs bosons at 14 TeV center-of-mass energies.\vspace*{4mm}\\

\section{Higgs boson production cross section in pp collisions}
As the main production mechanism of Higgs bosons is gluon fusion $pp\rightarrow gg\rightarrow H$, lowest order of the parton cross section $\sigma_{LO}(gg\rightarrow H)$ is expressed by the gluonic width of the Higgs boson 
$\Gamma_{LO}(H\rightarrow gg)$ \cite{6.}
\[\sigma_{LO}(gg\rightarrow H)=\frac{8\pi^2}{m^5_H}\Gamma_{LO}(H\rightarrow gg)\delta(\hat{s}-m^2_H)\ .\]
The lowest–order proton–proton cross section $\sigma_{LO}(pp\rightarrow H)$ can be defined by gluon luminosity as
\[\sigma_{LO}(pp\rightarrow H)=\sigma_0\tau_H\frac{d{\cal{L}}^{gg}}{d\tau_H}\ ,\]
where $s$ is the invariant pp collider energy squared, $\tau_H=\frac{m^2_H}{s}$ and
\[\sigma_0=\frac{G_F\alpha^2_s(\mu^2)}{288\sqrt{2}\pi}\Biggl|\frac{3}{4}
\sum\limits_QA_Q(\tau_Q)\Biggr|^2\ ,\]
where $G_F$ is Fermi coupling constant, $\alpha_s$ is strong coupling,
$A_Q$ denotes the quark amplitude and $\mu$ is the renormalization point and defines the scale parameter of $\alpha_s$.

	The $pp$ cross-section of Higgs bosons $\phi\in\{h, H, A\}$ formation at the NLO (next-to-leading order) QCD corrections, is written as follows:
\[\sigma(pp\rightarrow H + X)=\sigma_0\Biggl[1+C\frac{\alpha_s}{\pi}\Biggr]
\tau_H\frac{d{\cal{L}}^{gg}}{d\tau_H}+\]\[+\Delta\sigma_{gg}+\Delta\sigma_{gq}+\Delta\sigma_{q{\overline{q}}} \ ,\]
where $C$ arises from two-loop corrections of partonic cross-section, the quantities $\Delta\sigma_{gg}$, $\Delta\sigma_{gq}$ and $\Delta\sigma_{q{\overline{q}}}$ of the partonic cross section arise from $gg$, $gq$ and $q\overline{q}$ scattering. 

	The coupling of the Higgs $\phi$  to the bottom quarks in supersymmetric theory can be significant value comparable with the gluon-gluon fusion that is associated with large values of tan$\beta$. Accounting for the large tan$\beta$ values leads to the associated Higgs production $(b\overline{b})\phi + X$  illustrated in Fig. 4.\\
\bec
{\includegraphics[width=0.7\textwidth]{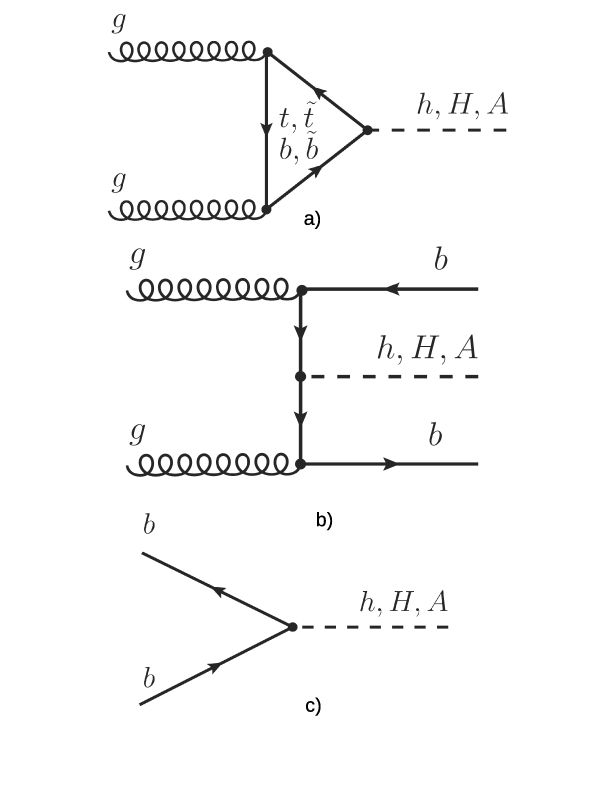}}\\
\emph{{Fig.4.}} {\emph{Leading order diagrams of the a) gluon fusion and b–associated Higgs production in the b) four-flavour and c) five-flavour scheme from \cite{5.}.}}\\
\ec

With the help of SusHi code (1.6.1 version) \cite{7.}, we carried out calculations of Higgs boson $H$ cross-section formation that include NNLO QCD contributions to LO quantities. This program also allows to calculate the differential cross sections of these processes with respect to the Higgs transverse momentum $p_t$ and (pseudo-)rapidity $y(\eta)$ through NNLO QCD contributions  \cite{8.}.
	The branching ratio of the Higgs boson for different benchmark scenarios, as well as the Yukawa coupling constants of the Higgs boson that are needed for the calculation of cross-sections were modelled using FeynHiggs code
(version 2.12.0) \cite{9.}. Minimal Supersymmetric Standard Model (MSSM) parameters were determined from the experimental data according to \cite{10.}, shown in Fig. 5 \vspace*{3mm} \\
\bec
{\includegraphics[width=0.61\textwidth]{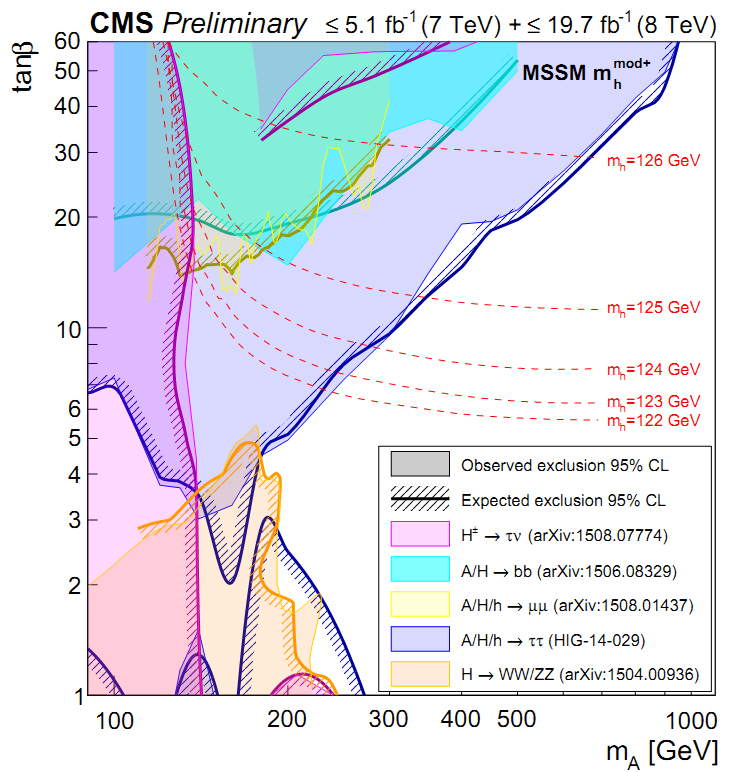}}\\
\emph{{Fig.5.}} {\emph{Restrictions on the parameters of the MSSM model from \cite{10.}.}}\\
\ec

From Fig. 5, it can be concluded that the decays of the Higgs boson via  fermionic channel are sensitive to large tangents, while the decays of the Higgs boson via bosonic channel are sensitive to the range of small tangents. This fact will be used by us during the cross section and branching ratio calculations of Higgs boson for an optimal agreement with the experimental data at low energies and for the best predictions at energies of 14 TeV at the LHC.

\section{Cross section and branching fraction calculations}
\subsection{The searches for a neutral Higgs boson, $H$ via fermion decay, $H\rightarrow\tau\tau$} 
Experimental data on the searches for the Higgs boson in the mass range 90-1000 GeV via decay channel $A\backslash H\backslash h\rightarrow\tau\tau$ are presented in \cite{5.}. The accuracy of the data calculations is based on the searches for three neutral Higgs bosons of MSSM model through the reconstruction of the invariant mass of two $\tau$ mesons with their subsequent decays into muons, electrons and hadrons, $\mu\mu, e\mu, \mu\tau_h, e\tau_h, \tau_h\tau_h$. To increase the accuracy of data analysis was measured cross section of b quark associated Higgs boson production, as an increase in constant coupling with $\tau$ leptons is observed for this process of Higgs boson creation. This channel of Higgs decay to $\tau\tau$ final state is perfect one to test the viability of MSSM model. Experimental data on the search for the Higgs boson for the gluon fusion process $(gg\phi)$ and the b–associated production process $(bb\phi)$ recorded by the CMS detector at 13 TeV centre–of–mass energy in 2015 are presented in Fig. 6 \vspace*{3mm}\\
\bec
{\includegraphics[width=0.5\textwidth]{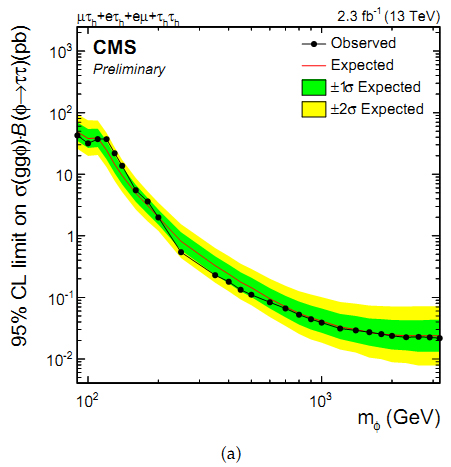}}\\
{\includegraphics[width=0.5\textwidth]{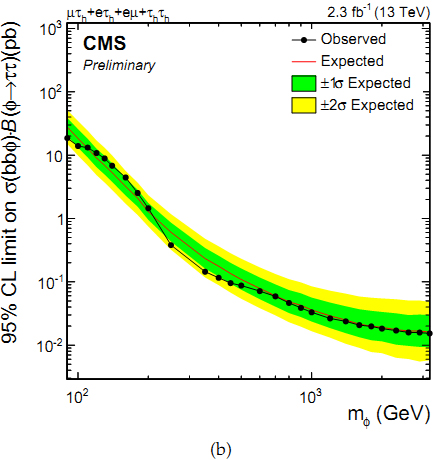}}\\
\emph{{Fig.6.}} {\emph{$\sigma(gg(bb)\rightarrow \phi)B(\phi\rightarrow\tau\tau)$ for a) the gluon fusion process $(gg\phi)$  and b) the b–associated production process $(bb\phi)$ from \cite{5.}.}}\\
\ec
Using the experimental data for the restriction of the numerical values of MSSM parameters, shown in Fig.6, with the help of computer program SusHi, we have calculated $\sigma(pp\rightarrow H)B(H\rightarrow\tau\tau (bb))$ for the gluon-gluon fusion $(gg\phi)$ and b–associated production process $(bb\phi)$. From the perspective of the searches for new physics at the LHC at an energy of 14 TeV, we have carried out calculations for energy of proton-proton collisions of 14 TeV, via two most probable decay channels of the Higgs boson $H\rightarrow\tau\tau$ and $H\rightarrow bb$ for the gluon-gluon or b-associated process of the Higgs boson formation. The results of our calculations are presented in Fig. 7\\
\bec
{\includegraphics[width=0.7\textwidth]{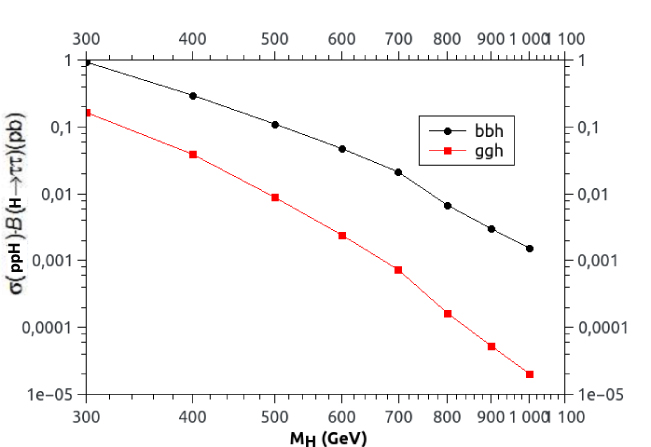}}\\
{\includegraphics[width=0.7\textwidth]{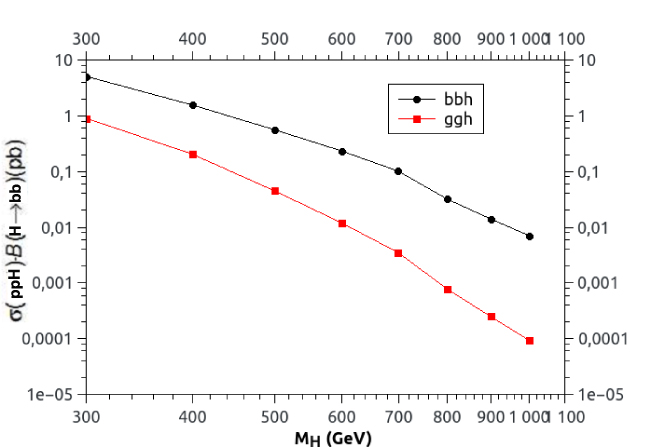}}\\
\emph{{Fig.7.}} {\emph{$\sigma(pp\rightarrow H)B(H\rightarrow\tau\tau (bb))$ for the gluon-gluon fusion and b–associated production process in the range of  Higgs mass $M_H$ = 300 - 1000 GeV with tan$\beta$ = 17 and energy 14 TeV for two decay channels of the Higgs boson, $H\rightarrow\tau\tau$ (up) and $H\rightarrow bb$ (down).}}\\
\ec
From Fig. 7 it can be seen the increase (by one order of magnitude) of $\sigma(ppH)B(H\rightarrow bb)\ (pb)$ compared with the value $\sigma(ppH)B(H \rightarrow\tau\tau)$ that emphasizes the importance of accounting of other neutral Higgs boson decay channels at the LHC, in particular, $H\rightarrow bb$. In addition, we see a significant predominance of $(bb\phi)$  Higgs production process compared with the process $(gg\phi)$  that confirms the theoretical predictions of the prevalence of this  process due to increase of Higgs boson Yukawa coupling constant for large values of tan$\beta$.\\
\hspace*{6mm}	To study the kinematics of the processes, we have calculated differential cross sections with respect to the Higgs transverse momentum $p_t$ and pseudorapidity, $\eta$ at 14 TeV, that are presented in Fig. 8\\
\bec
{\includegraphics[width=0.7\textwidth]{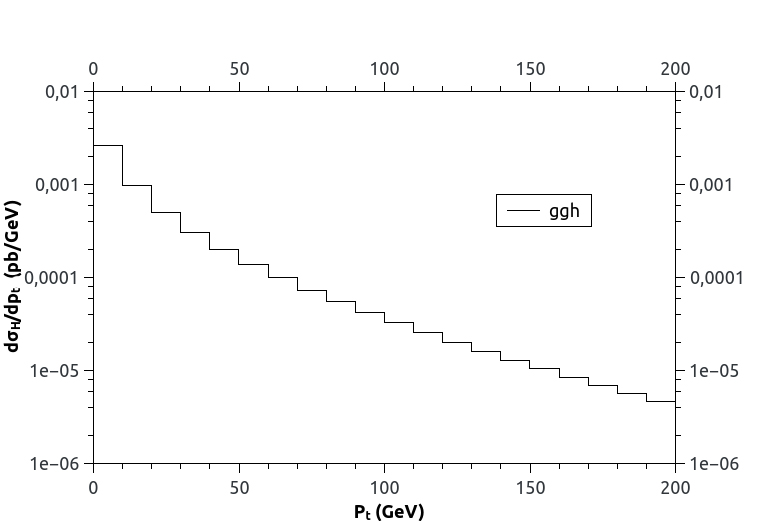}}\\
{\includegraphics[width=0.7\textwidth]{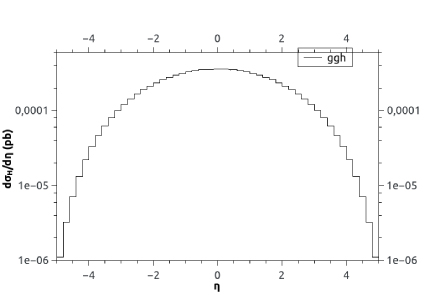}}\\
\emph{{Fig.8.}} {\emph{Differential cross sections with respect to the Higgs boson $H$ transverse momentum $p_t$, (up) and pseudorapidity, $\eta$ (down) at the energy of 14 TeV.}}\\
\ec
From Fig.8 is seen that differential cross section smoothly decreases for $ggh$ process. The character of the differential cross section with respect to the pseudorapidity indicates that the process of Higgs boson decay is accompanied by the direction of decay products that are perpendicular to the axis of the of the proton-proton collisions. This process is also characterized by a large value of the differential cross section in the region of pseudorapidity, $\eta$ = 1-2.2, that corresponds to the angles relative to the collision axis of $\sim 50^0 – 10^0$.

\subsection{The searches for a pseudoscalar boson, $A$ via dibozon decay, $A \rightarrow Zh$}

Experimental data on the searches for a pseudoscalar boson $A$ in the mass range of 200-600 GeV decaying into a $Z$ boson and the SM-like Higgs boson $h$, where $h$ boson decays into a pair $bb$ and $Z$ boson decays into a pair of oppositely-charged electrons or muons, were presented in \cite{11.}. The data from proton-proton collisions at a center-of-mass energy 8 TeV collected with the CMS detector correspond to an integrated luminosity of 19.7 fb$^{-1}$. $A$ boson is produced via the gluon-gluon fusion and its branching fraction into $Zh$ is relatively large compared to other channels. Furthermore, this channel is selected because of the lightness of detection of $Z$ and $h$ decay products, $Z\rightarrow ll$ and $h\rightarrow bb$. Branching fractions for these decay channels are large values all over parameter space of 2HDM model. The upper limit on the $\sigma_A B(A\rightarrow Zh\rightarrow llbb)$ , in the mass range of $A$ boson $M_A$ = 200-600 GeV is presented in Fig. 9 \\
\bec
{\includegraphics[width=0.7\textwidth]{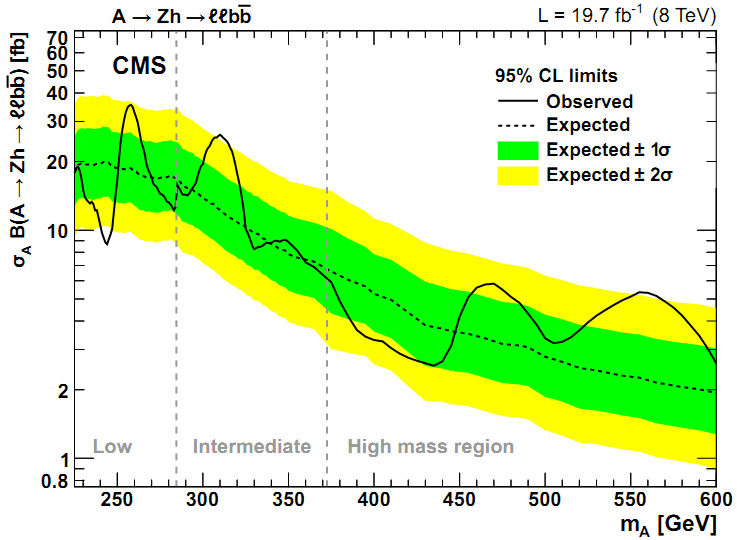}}\\
\emph{{Fig.9.}} {\emph{Observed and expected 95$\%$ CL upper limit on $\sigma_A B(A\rightarrow Zh\rightarrow llbb)$ as a function of $M_A$ from \cite{11.}.}}\\
\ec
Since in this mass range exists a peak with a local significance of 2.6$\sigma$ or a global 1.1$\sigma$ significance, it would be interesting to check its presence at higher energies and luminosities. With the help of the program SusHi we have calculated $\sigma_A B(A\rightarrow Zh\rightarrow llbb)$ in the mass range $M_A$ = 300-1000 GeV with tan$\beta$ = 2. The results of our calculations at a center-of-mass energy of 8 TeV and at the projected at the LHC energies of 14 TeV are presented in Fig. 10\\
\bec
{\includegraphics[width=0.7\textwidth]{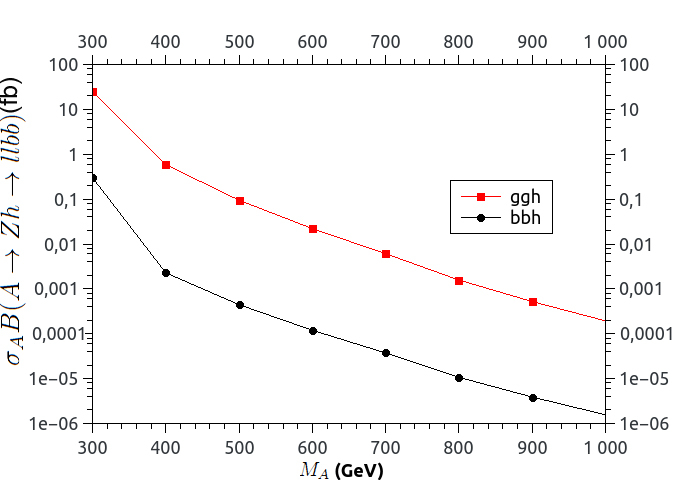}}\\
{\includegraphics[width=0.7\textwidth]{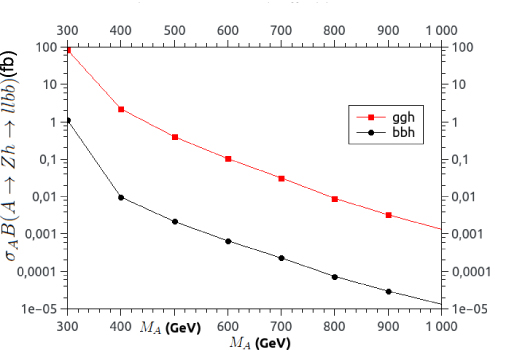}}\\
\emph{{Fig.10.}} {\emph{$\sigma_A B(A\rightarrow Zh\rightarrow llbb)$ as a function of  $M_A$  at a center-of-mass energy of 8 TeV (up) and of 14 TeV (down).}}\\
\ec
From Fig. 14 is seen an increase (by one order of magnitude) of the value $\sigma_A B(A\rightarrow Zh\rightarrow llbb)$  at $M_A$ = 1000 GeV, that allows us to assume optimistically the possible discovery of a pseudoscalar boson, $A$ at higher energies and luminosities at the LHC. In addition, we see the predominance of $A$ boson production through gluon-gluon fusion, that emphasizes the correctness of the theoretical predictions with respect to the substantial Higgs interaction with the $b$ quarks only at high tan$\beta$.

	The calculations of differential cross sections for pseudoscalar boson with respect to the transverse momentum $p_t$ and pseudorapidity $\eta$ at a center-of-mass energy of 14 TeV are presented in Fig. 11 \vspace*{3mm}\\
\bec
{\includegraphics[width=0.7\textwidth]{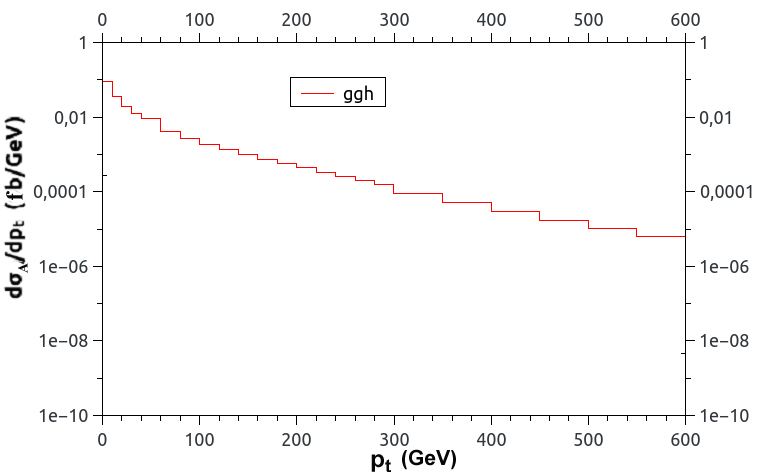}}\\
{\includegraphics[width=0.7\textwidth]{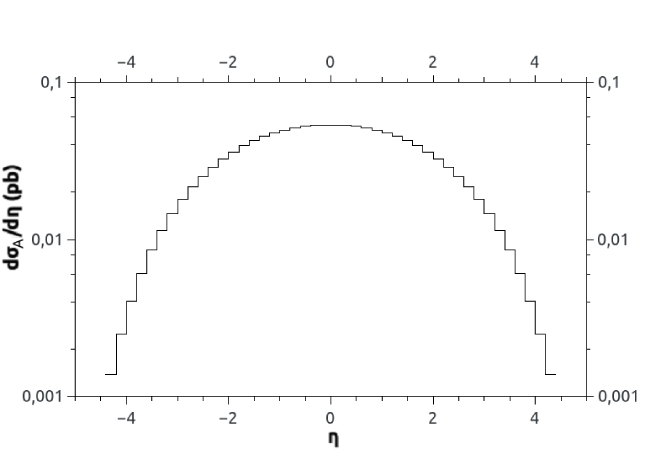}}\\
\emph{{Fig.11.}} {\emph{Differential cross sections for pseudoscalar boson, $A$ with respect to the transverse momentum $p_t$ (up) and pseudorapidity $\eta$ (down) at a center-of-mass energy of 14 TeV.}}\\
\ec
It should be noted that the differential cross section with respect to the transverse momentum $p_t$, smoothly decreases for $ggh$ process and is maximal for small values of transverse momentum. The character of the differential cross section with respect to the pseudorapidity indicates that the process of Higgs boson decay does not have a preferred direction perpendicular to the proton-proton collision axis, that emphasizes the importance of searches for the Higgs boson $A$ in all directions with respect to the collision axis. From Fig.11 is seen the significant predominance of the value of the differential cross sections with respect to the pseudorapidity for process $A\rightarrow Zh\rightarrow llbb$ compared to the data for $H\rightarrow\tau\tau)$ process of the Higgs boson formation, presented in Fig. 8.

\subsection{The searches for a heavy scalar boson, $H$ via dibozon decay, $H \rightarrow hh$}

Due to the large amount of data on decay channels, there was selected the decay channel, $H\rightarrow hh(bb\tau\tau)$, predicted in the MSSM model. We have performed calculations of cross sections of Higgs boson formation using experimental data with three final states, $e\tau_h, \mu\tau_h, \tau_h\tau_h$, where $\tau_h$ - a $\tau$ lepton decaying into hadrons \cite{12.}. Parameter space is selected for largest cross section values and the range of $M_A$ is selected with respect to recent experimental data, presented in Fig. 12.\\
\bec
{\includegraphics[width=0.7\textwidth]{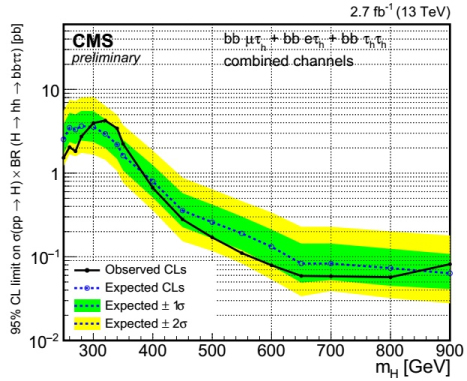}}\\
\emph{{Fig.12.}} {\emph{The upper limit on $\sigma(pp\rightarrow H)\times BR(H\rightarrow hh \rightarrow bb \tau\tau)$ as the function of $m_H$ from \cite{12.}.}}\\
\ec

In the experiment was studied the resonant Higgs boson production via the process $pp\rightarrow H\rightarrow hh\rightarrow bb\tau\tau$, where $H$ is the CP-even Higgs boson of unknown mass. The branching ratio, $H\rightarrow hh$ can be large for small values of tan$\beta$ caused by the experimentally measured value of Higgs boson mass, $m_h\simeq 125$ GeV. The searches for the final state $bb\tau\tau$ are carried out taking into account the most probable decay channels of $\tau$ leptons: $e\tau_h, \mu\tau_h, \tau_h\tau_h$. Fig.12 presents the upper limit on the $\sigma(pp\rightarrow H)\times BR(H\rightarrow hh \rightarrow bb \tau\tau)$ for the combination of the three channels as a function of the resonance mass $m_H$.

	Using the experimental data presented in Fig.12, with the help of the computer program SusHi, we calculated $\sigma(pp\rightarrow H)\times BR(H\rightarrow hh\rightarrow bb\tau\tau)$ for the gluon-gluon fusion $(gg\phi)$ and b-associated production process  $(bb\phi)$   at a centre-of-mass energy of 14 TeV and tan$\beta$ = 2 in the mass range $M_H$ = 300 - 1000 GeV, Fig. 13\\
\bec
{\includegraphics[width=0.7\textwidth]{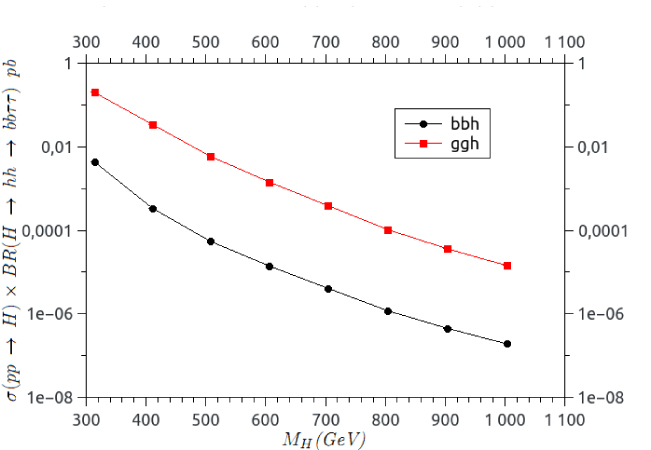}}\\
\emph{{Fig.13.}} {\emph{$\sigma(pp\rightarrow H)\times BR(H\rightarrow hh\rightarrow bb\tau\tau)$ for the gluon-gluon fusion 
$(gg\phi)$ and b-associated production process $(bb\phi)$ at a centre-of-mass energy of 14 TeV at the LHC.}}\\
\ec

In addition, it should be noted the predominance of the process of the gluon-gluon fusion of the Higgs boson production, compared with b-associated production process in the region of small values of tan$\beta$ that is differ from analogous calculations for large values of tan$\beta$, where dominated $bbh$ processes, presented in Fig.7.

		The kinematics of the process $pp\rightarrow H\rightarrow hh\rightarrow bb\tau\tau$  is presented by calculations of differential cross sections with respect to the Higgs transverse momentum $p_t$ and pseudorapidity $\eta$ at 14 TeV, Fig. 14\\
\bec
{\includegraphics[width=0.7\textwidth]{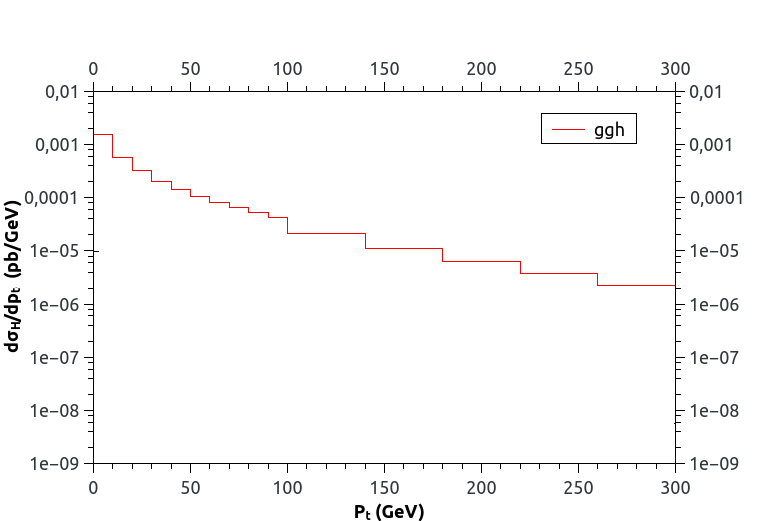}}\\
{\includegraphics[width=0.7\textwidth]{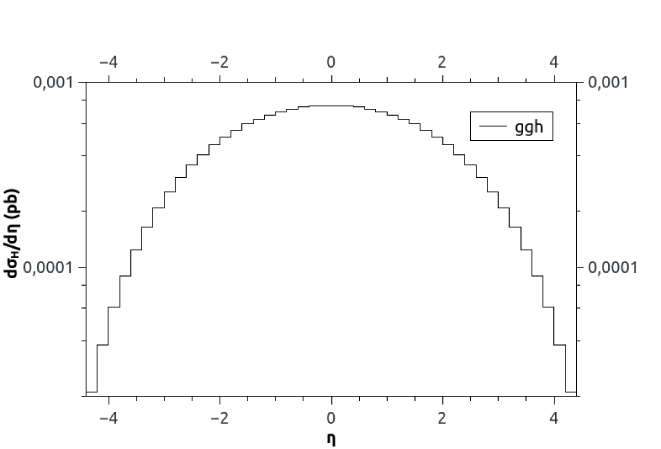}}\\
\emph{{Fig.14.}} {\emph{Differential cross sections for Higgs boson, $H$ with respect to the transverse momentum $p_t$ (up) and pseudorapidity $\eta$ (down) at a center-of-mass energy of 14 TeV.}}\\
\ec

The character of the differential cross section does not differ from the previous cases. This fact underscores the dependence of this characteristic from many others factors beyond the parameter space data. Blurred peak also indicates the large range of emission angles of the decay products of the Higgs boson with respect to the axis of the proton-proton collisions. 

\section{Conclusion}
We have calculated the production of cross section on branching fraction, $\sigma\times Br$ for two mechanisms of production and three decay mechanisms of Higgs bosons within 2HDM model: 1)$pp\rightarrow H\rightarrow\tau\tau$, 2)$pp\rightarrow A\rightarrow Zh \rightarrow llb\overline{b}$, 3) $pp\rightarrow H\rightarrow hh\rightarrow bb\tau\tau$. With the help of a computer program SusHi were carried out calculations for 8 TeV, as well as for the projected at the LHC energies of 14 TeV in the center-of-mass energy. The obtained calculations present an increase in the value of $\sigma\times Br$ for three considered decay processes of the Higgs boson, but in the third case this increase is insignificant. In all three cases, were compared cross sections of  the Higgs boson production via gluon-gluon fusion and b-associated production process and found the predominance of the $(bb\phi)$ process only for fermionic decay channel of the Higgs boson, $H\rightarrow \tau\tau$, when the value of tan$\beta$ was significant one. In the other two bozonic decay channels there was a significant excess of the cross-section of the Higgs boson production via gluon-gluon fusion for small values of tan$\beta$. For three considered cases are calculated the differential cross sections for Higgs boson with respect to the transverse momentum $p_t$ and pseudorapidity $\eta$ at a center-of-mass energy of 14 TeV. The distribution of the differential cross section with respect to pseudorapidity does not detect certain direction of decay products to the axis of the proton-proton collisions.

\label{page-last} 
\label{last-page}
\end{document}